\documentclass[conference]{IEEEtran}
\IEEEoverridecommandlockouts
\usepackage[utf8]{inputenc}
\usepackage[english]{babel}
\usepackage{cite}
\usepackage{amsmath,amssymb,amsfonts}
\usepackage{algorithmic}
\usepackage{graphicx}
\usepackage{textcomp}
\usepackage{caption}
\usepackage{subcaption}
\usepackage{multirow}
\usepackage{booktabs}
\usepackage[draft=true]{minted}
\usepackage[breaklinks=true]{hyperref}
\usepackage{breakcites}
\usepackage[dvipsnames]{xcolor}

\setminted{fontsize=\footnotesize,
    xleftmargin=8mm,
    linenos,
    frame=lines,
    breaklines}

\makeatletter

\def\BibTeX{{\rm B\kern-.05em{\sc i\kern-.025em b}\kern-.08em
    T\kern-.1667em\lower.7ex\hbox{E}\kern-.125emX}}
    
\makeatletter

\begin{document}

\title{An Empirical Analysis of UI-based Flaky Tests}

\author{\IEEEauthorblockN{Alan Romano$^{1}$,
Zihe Song$^{2}$, Sampath Grandhi$^{2}$, Wei Yang$^{2}$, and Weihang Wang$^{1}$}
\IEEEauthorblockA{$^{1}$\textit{University at Buffalo, SUNY} $^{2}$\textit{University of Texas at Dallas}}
}
\maketitle

\begin{abstract}
Flaky tests have gained attention from the research community in recent years and with good reason. These tests lead to wasted time and resources, and they reduce the reliability of the test suites and build systems they affect. However, most of the existing work on flaky tests focus exclusively on traditional unit tests. This work ignores UI tests that have larger input spaces and more diverse running conditions than traditional unit tests. In addition, UI tests tend to be more complex and resource-heavy, making them unsuited for detection techniques involving rerunning test suites multiple times.

In this paper, we perform a study on flaky UI tests. We analyze 235 flaky UI test samples found in 62 projects from both web and Android environments. We identify the common underlying root causes of flakiness in the UI tests, the strategies used to manifest the flaky behavior, and the fixing strategies used to remedy flaky UI tests. The findings made in this work can provide a foundation for the development of detection and prevention techniques for flakiness arising in UI tests.
\end{abstract}

\section{Introduction}
Software testing is a significant part of software development.  Most developers  write test suites to repeatedly test various indicators of functioning software. If a test fails, developers will analyze the corresponding test to debug and fix the software. However, not all testing results are fully reliable. Sometimes the test may show flakiness, and a test showing this behavior is denoted as a flaky test.

Flaky tests refer to tests with unstable test results. That is, the same test suite sometimes passes and sometimes fails under the exact same software code and testing code. The existence of flaky tests destroys the deterministic relationship between test results and code quality. Once a flaky test appears, it may lead to tons of efforts wasted in debugging the failed test, which leads to delays in software release cycle and reduced developer productivity~\cite{micco2017state}.

In the past few years, researchers have increased efforts to address this problem~\cite{luo_2014,bell2018deflaker,lam2019root}. 
However, the existing research on flaky tests mostly focuses on unit tests. Compared with traditional unit testing, the execution environment and automation process of UI testing are significantly different: 
First, many of the events in these tests, such as  handling user input,
making operating system (or browser) API calls, and downloading and rendering multiple resources (such as images and scripts) required by the interface, are highly asynchronous in nature. This means that user events and various other tasks will be triggered in a non-deterministic order. 

Second, compared to traditional unit testing, flaky UI tests are more difficult to detect and reproduce. This is because it is difficult to cover all use-case scenarios by simulating user events to automate UI testing. UI tests introduce new sources of flakiness, either from the layer between the user and the UI or the layer between the UI and the test/application code. Moreover, the execution speed of the UI test in continuous integration environments is slow, and this difference in execution speed makes detecting and reproducing flaky tests more difficult. 
Therefore, researching flaky UI tests can help web and mobile UI developers by providing insights on effective detection and prevention methods.

To further investigate flaky UI tests, we collect and analyze 235 real-world flaky UI test examples found in popular web and Android mobile projects. For each flaky test example, we inspect commit descriptions, issue reports, reported causes, and changed code.
We focus on the following questions and summarize our findings and implications in Table~\ref{tab:findings}.

{\bf RQ1: What are the typical root causes behind flaky UI tests?} We examine the collected flaky UI test samples to determine the underlying causes of flaky behavior. We group similar root causes together into 4 main categories: {\it Async Wait}, {\it Environment}, {\it Test Runner API Issues}, and {\it Test Script Logic Issues}. 

{\bf RQ2: What conditions do flaky UI tests manifest in and how are they reproduced?} In order to understand how users report intermittent behaviors, we investigate the common strategies used to manifest the flaky UI test samples. The data reveals 5 strategies used to reproduce and report flaky UI test behavior: {\it Specify Problematic Platform}, {\it Reorder/Prune Test Suite}, {\it Reset Configuration Between Tests}, {\it Provide Code Snippet}, and {\it Force Environment Conditions}.

{\bf RQ3: How are these flaky UI tests typically fixed?} We identify the bug fix applied to each collected flaky UI test sample and group similar ones together. We find 4 main categories for bug fixing strategies: {\it Delay}, {\it Dependency}, {\it Refactor Test}, and {\it Disable Features}. 

We investigate the impacts that these UI-specific features have on flakiness, and we find several distinctions. Based on the investigation of above research questions, the main contributions of this study are:
\begin{enumerate}
    \item Our study provides guidance for developers to create reliable and stable UI test suites, which can reduce the occurrence of flaky UI tests.
    \item Our study summarizes the commonly-used manifestation and fix strategies of flaky UI tests to help developers easily reproduce and fix flaky tests, allowing them to avoid wasted development resources.
    \item Our study motivates future work for automated detection and fixing techniques of UI-based flaky tests.
\end{enumerate}

\begin{table*}
\caption{Summary of Findings and Implications}
\label{tab:findings}
\resizebox{\textwidth}{!}{%
\begin{tabular}{|l|l|l|}
\hline
  & \textbf{Findings}  & \textbf{Implications}  \\ \hline
1 & \begin{tabular}[c]{@{}l@{}}Of the observed flaky tests collected, 105 tests of the 235 (45.1\%) \\ dataset are caused by an Async Wait issue.\end{tabular}        & \begin{tabular}[c]{@{}l@{}}This group represents a significant portion of the dataset collected and highlights\\ the need to take this root cause into consideration when designing UI tests.\end{tabular} \\ \hline

2 & \begin{tabular}[c]{@{}l@{}}Async Wait issues are more prevalent in web projects rather than mobile\\ projects (W 52.0 \% vs M 32.5\%).\end{tabular}              & \begin{tabular}[c]{@{}l@{}}The web presents an environment with less stable timing and scheduling\\ compared with a mobile environment, so more care must be taken when\\ network or resource loads are used within web UI tests.\end{tabular} \\ \hline

3 & \begin{tabular}[c]{@{}l@{}}Platform issues are happening more frequent on mobile projects rather\\ than web projects (W 10.5 \% vs M 21.7\%).\end{tabular} & \begin{tabular}[c]{@{}l@{}}It may be caused by Android fragmentation problem. So the Android\\ developers should pay more attention to the environment configuration\\ when choosing the test model.\end{tabular} \\ \hline

4 & \begin{tabular}[c]{@{}l@{}}Layout difference (cross-platform) root causes are found more in web\\ flaky test than in mobile flaky tests (W 5.3 \% vs M 1.2\%).\end{tabular}                                                             & \begin{tabular}[c]{@{}l@{}}This difference can be explained by the number of additional platform conditions that\\ web applications can be exposed compared with the conditions found in \\ mobile environment, such as different window sizes, different browser \\ rendering strategies, etc...\end{tabular} \\ \hline

5 & \begin{tabular}[c]{@{}l@{}}Besides removing flaky test, the most common fixing strategies are\\ refactoring logic implementations (46.0\%) and fixing delays (39.3\%).\\ Among them, refactoring logic implementations can solve most issues\\ caused by wrong test script logic, and fixing delay strategy can solve \\ most timing issues.\end{tabular} 
& \begin{tabular}[c]{@{}l@{}}Refactoring logic implementations and fixing delays should be the first-\\ considered strategies for developers when fixing bugs.\end{tabular} \\ \hline

6 & \begin{tabular}[c]{@{}l@{}}Dependency fixes are more common in mobile projects than web \\ projects (W 1.3\% vs M 21.4\%).\end{tabular}                & \begin{tabular}[c]{@{}l@{}}This trend can be caused by the Android fragmentation problem. Android developers \\ should pay more attention to this problem when designing test suites.\end{tabular} \\ \hline

7 & \begin{tabular}[c]{@{}l@{}}Delay fixes are more common in web projects than mobile projects\\  (W 32.2\% vs M 17.9\%).\end{tabular}                          & \begin{tabular}[c]{@{}l@{}}This phenomenon is related to the most common test framework in Android \\ testing, Espresso, which recommends disabling animations during tests.\end{tabular} \\ \hline
\end{tabular}%
}
\end{table*}

\section{Background}

\subsection{Impacts of Flaky UI Tests}

\subsubsection{Individual Test Failures}
The simplest impact that flaky test can have on test suites is that the individual test run will fail. This flaky behavior leads to a minimal amount of time and resources wasted by attempting to retry the single test. 

\subsubsection{Build Failures}
Flaky tests that are part of continuous integration systems can lead to intermittent build failures. Flaky tests in this stage lead to wasted time trying to identify the underlying cause of the build failure only to find out that the failure was not caused by a regression in the code. 

\subsubsection{CI Test Timeouts}
Some flaky behaviors do not cause the tests to fail outright. Instead, they cause hangups in the CI system that lead to timeouts. These hangups waste time as the system waits for a process that never finishes, causing the CI system to wait until a specified timeout is met.

\section{Methodology}
\subsection{Sample Collection}
\subsubsection{Web}
\begin{table}
\caption{Summary of Commit Info from UI Frameworks
}
\label{table:commits_info}
\resizebox{1\columnwidth}{!}{
\begin{tabular}{@{}lrrrr@{}}
\toprule
{\bf UI Topic} & 
{\bf Projects} & {\bf Commits} &
\textbf{\begin{tabular}[c]{@{}r@{}}Flaky Keyword \\ Filtering\end{tabular}} &
\textbf{\begin{tabular}[c]{@{}r@{}}UI Keyword \\ Filtering\end{tabular}}
\\
\hline
web	      & 999		 & 	772,901	  & 2,553	 & 210	 \\
angular	  & 998		 & 	407,434	  & 222		 & 19	 \\
vue	      & 998		 & 	344,526	  & 52		 & 1	 \\
react	  & 997		 &  1,110,993 & 603		 & 30	 \\
svg	      & 995		 & 	135,563	  & 24		 & 1	 \\
bootstrap & 995		 & 	98,264	  & 112		 & 0	 \\
d3	      & 980		 & 	106,160	  & 82		 & 1	 \\
emberjs	  & 629		 & 	3,961	  & 1		 & 0	 \\
\hline 
Total & 7,590 & 2,979,802 & 3,649 & 262\\
Distinct & 7,037 & 2,613,420 &3,516 & 254\\
\bottomrule
\end{tabular}
}
\end{table}

In order to collect samples of flaky UI tests, we retrieve commit samples from GitHub repositories. 
First, we obtain a list of the repositories leveraging popular web UI frameworks using the topic keywords `react', `angular', `vue', `emberjs', `d3', `svg', `web', and 'bootstrap'. These keywords are used with the GitHub Search API~\cite{github-search-api} to identify 7,037 distinct repositories pertaining to these topics. From this set of repositories, we download all of the commits for these projects giving a total of 2,613,420 commits to search. Next, we follow a procedure similar to the one used in Luo {\it et al.} ~\cite{luo_2014} and search the commit messages for the patterns `flak*' and `intermit*' and the word `test' to find commits marked as flaky. This step reduces the number of commits to 3,516. In order to confirm which of these commits were flaky UI tests, manual inspection was performed on the commits in the list. In order to expedite the process, another keyword search is performed using the keywords `ui', `gui', `visual', `window', `button', `display', `click', and 'animation' to prioritize the commits most likely to refer to flaky UI tests. This final search prioritizes 254 commit messages to search, but the full 3,516 are searched to increase the chance of identifying flaky UI test commits. After manual inspection and removing duplicate commits, the number of verified flaky tests is 152. Table~\ref{table:commits_info} shows the summary of commit information.

\subsubsection{Android}
Compared with web development, Android developers are not as consistent with their choice of UI framework. Therefore, to find flaky UI tests on the Android platform, we use the GitHub Archive\cite{gharchive} to perform a massive search on all commits and issue reports on GitHub instead of focusing on repositories using popular UI framework. Specifically, We limit our search to the list of closed issues, since the flaky tests in closed issues are more likely to be fixed than the tests in open issues. We search with the keywords `flak*', `intermit*', and `test*', which is similar to the patterns used in web searching. In order to ensure the issues we find are flaky UI tests on the Android platform, we also add constraints like `android', `ui', `espresso', `screen', etc.

\subsection{Portion of Flaky UI Tests to Other Tests}
We find that flaky UI tests collected in our methodology make up a small portion of all tests available in these repositories. This small portion can be explained by several reasons. Based on GH Archive~\cite{gharchive}, the number of open issues (over 70,000) containing potential flaky UI tests outnumbers those in closed issues (over 30,000). Open issues cannot be included in our study; however, this large number of open issues possibly containing flaky UI tests highlights the significance of UI flakiness. Besides, there are flaky UI tests not captured through the keywords. One example is in the {\tt material-components-web} repository [2]. While our dataset is not exhaustive, we believe the results can provide a basis for future work to build on.

\subsection{Sample Inspection}
\begin{table}
\caption{Top 10 Projects Containing the Most Flaky Tests}
\label{table:proj_stats}
\resizebox{1\columnwidth}{!}{
\begin{tabular}{@{}lrrr@{}}
\toprule
{\bf Project} & 
\textbf{\begin{tabular}[c]{@{}l@{}}Inspected \\ Commits\end{tabular}}
& {\bf Flaky Tests} & {\bf LOC}\\
\hline
Waterfox	 &	937	 & 	23	 & 	3,949,098	 \\ 
qutebrowser	 &	124	 & 	4	 & 	45,313	 \\ 
influxdb	 &	81	 & 	12	 & 	124,591	 \\ 
angular	 &	69	 & 	2	 & 	135,253	 \\ 
plotly.js	 &	37	 & 	24	 & 	760,504	 \\ 
material-components-web	 &	26	 & 	5	 & 	78,972	 \\ 
components	 &	21	 & 	5	 & 	34,715	 \\ 
oppia	 &	20	 & 	2	 & 	132,284	 \\ 
wix-style-react	 &	15	 & 	11	 & 	19,830	 \\ 
streamlabs-obs	 &	13	 & 	1	 & 	179,184	 \\ 
\hline 
material-components-android &  358		 & 9		 & 11,682 \\
Focus-android   & 24  & 8		 & 30,952	 \\
RxBinding	    & 21  & 6	& 8,787	 \\
Xamarin.Forms   & 140  & 5  & 46,609	 \\
FirebaseUI-Android	& 34 & 4  & 4,386	 \\
Fenix	        & 20   & 4		 & 155,50 	 \\
Detox           & 38 & 3        & 2,254 \\
Components 	    & 14  & 3		 & 34,715	 \\
Mapbox-navigation-android   & 4 & 2 & 6,201	 \\
Sunflower       & 3  & 1		 & 354	 \\
\bottomrule
\end{tabular}
}
\end{table}

After collecting these commits and issues of flaky tests reports from GitHub, we manually inspect the collected samples to identify the information relevant to our research questions. In particular, we analyze the collected flaky tests by first inspecting the commits in the web projects and the issue reports in the Android projects for the following traits: the root cause of the flakiness, how the flakiness is manifested, how the flakiness was fixed, the test affected, the testing environment, and the lines of code of the fix. For the commits, we inspect the commit message, changed code, and linked issues. For the issue reports, we inspect the developer comments and the linked commits. When available, we also inspect the execution logs from the CI.
Table~\ref{table:proj_stats} shows the information of projects containing flaky tests. Through inspection, we obtained the sample set of 235 flaky tests, of which 152 were from web repositories and 83 were from Android repositories.

\subsection{Dataset Composition}
Our dataset consists of a diverse set of flaky UI test samples. The languages of the flaky UI tests analyzed are JavaScript (63.8\%), TypeScript (20.4\%), HTML (8.6\%), and others (7.2\%) for the web projects and Java (48.2\%), Kotlin (21.7\%), and others (30.1\%) for the Android projects.

\section{Cause of Flakiness}
\label{sec:root-cause}
We investigate the collected flaky tests to determine the root cause of the flaky behavior. We manually inspect the related commits and issues of the test in order to locate the code or condition that caused the flakiness. 
We base our root cause categories on those defined by Luo et al.~\cite{luo_2014}. We extend the set of categories to include new categories specific to UI flakiness (“Animation Timing Issue”, “DOM Selector Issue”, etc...). 
The categorization results are summarized in Table~\ref{table:root_cause_category_stats}. 

\begin{table}[hbpt]
\caption{Summary of Root Cause Categories Found}
\label{table:root_cause_category_stats}

\resizebox{1\columnwidth}{!}{
    \begin{tabular}{@{}llrrr@{}}
    \toprule
    \begin{tabular}[c]{@{}l@{}}{\bf Root Cause}\\ {\bf Categories}\end{tabular} & \begin{tabular}[c]{@{}l@{}}{\bf Root Cause}\\ {\bf Subcategories}\end{tabular} & {\bf Web} & {\bf Mobile} & {\bf Total} \\ \midrule
       Async Wait      & Network Resource Loading   & 15  & 4   & 19 \\
       & Resource Rendering                & 47  & 14  & 61 \\
            & Animation Timing Issue            & 17  & 9   & 26 \\ \hline
Environment & Platform Issue                    & 16  & 18  & 34 \\
            & Layout Difference                 & 9   &  1  & 10 \\ \hline
Test Runner & DOM Selector Issue                & 13  &  3  & 16 \\
API Issue   & Incorrect Test Runner Interaction & 10  & 14  & 24 \\ \hline
Test Script & Unordered Collections             & 5   & 0   & 5 \\
Logic Issue & Time                              & 1  & 0  & 1\\
            & Incorrect Resource Load Order     & 11  & 11  & 22\\
            & Test Order Dependency             & 6   & 6   & 12 \\
            & Randomness                        & 2   & 3   & 5 \\ \hline
            & Total                             & 152 & 83  & 235 \\ \bottomrule
    
    \end{tabular}
}
\end{table}

\subsection{Categorization}
After manual inspection of the flaky UI tests,
we identify four categories that the root causes of flakiness in these tests can fall under: (1) Timing Issue, (2) Platform Issue, (3) Test Runner API Issue, and (4) Test Script Logic Issue. We describe the categories and provide examples for each below.

\subsubsection{Async Wait}
We have found the root cause for a significant portion (45\%) of the flaky tests analyzed arise from issues with async wait mechanisms. The common cause of such issues is that the tests do not properly schedule fetching and performing actions on program objects or UI elements, causing issues with attempting to interact with elements that have not been loaded completely. The program objects or UI elements can come from network requests, the browser rendering pipeline, or graphics pipeline. This improper ordering of events results in an invalid action and causes an exception to be thrown. Among these async wait issues, we identified three three subcategories that group similar root causes together. 

\paragraph {Network Resource Loading}
Flaky tests in this category attempt to manipulate data or other resources before they are fully loaded. Attempting to manipulate nonexistent data causes exceptions in the test. An example is seen in the {\tt ring-ui}~\cite{ring-ui-rc-eg} web component library repository. This library provides custom-branded reusable web components to build a consistent theme across web pages. In this project, some components being tested utilize images that are fetched through network requests; however, depending on the network conditions, the images may fail to load on time or fail to load altogether. The code snippet in Figure~\ref{fig:ring-ui-rc-eg} shows how the \verb|url| variable defined in Line 1 is URL for an image network call to an external web server. The image is an avatar used by the {\tt tag} component on Line 8 to display on the page. When the server call occasionally fails to respond in time due to a heavy network load, the visual test will 
intermittently fail as the rendered {\tt tag} component will be missing the image. 

\begin{figure}[ht]
\begin{minted}
[highlightlines={1}]
{js}
const url = `${hubConFigureserverUri}/api/rest/avatar/ default?username=Jet%20Brains`;

class TagDemo extends React.Component {
render() {
  return (
    <div>
      <Tag>Simple</Tag>
      <Tag avatar={url} readOnly={false}>
        With avatar
      </Tag>
    </div>
  );
}}
\end{minted}
\caption{{\tt ring-ui} Network Resource Loading Example.}
\label{fig:ring-ui-rc-eg}
\end{figure}

\begin{figure}
    \centering
    
    \begin{subfigure}{\columnwidth}
        \includegraphics
        [width=\textwidth]
        {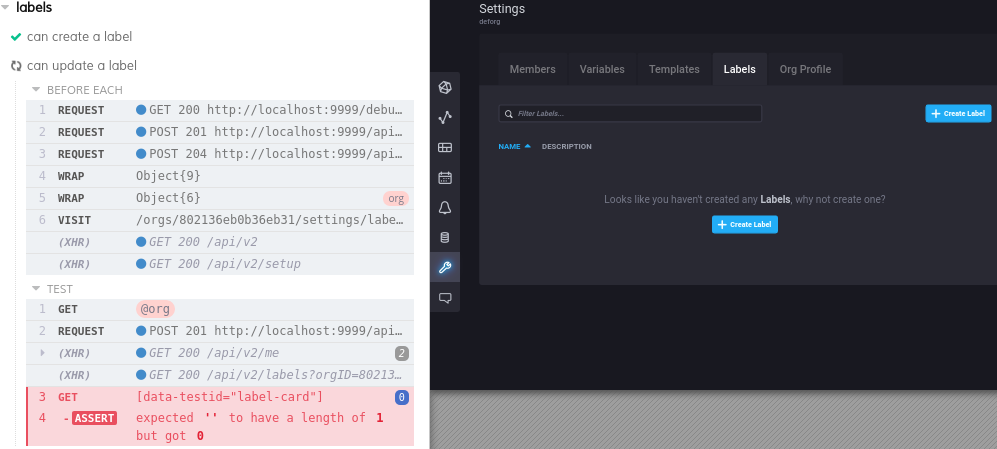}
        \caption{The test for updating a label first creates a new label through the UI. In this case, the backend had not finished processing the new label, so the network call to fetch all labels returns an empty response.}
        \label{subfig:influxdb_screenshot_rc}
    \end{subfigure}
    
    \begin{subfigure}{\columnwidth}
    \begin{minted}
    [highlightlines={6-11,13-17}]
    {js}
it('can update a label', () => {
  ...
  const newLabelName = 'attribut ()'
  const newLabelDescription = "..."

  // create label
  cy.get<Organization>('@org').then(({id}) => {
    cy.createLabel(oldLabelName, id, {
      description: oldLabelDescription,
    })
  })

  // verify name, descr, color 
  cy.getByTestID('label-card')
    .should('have.length', 1)
  cy.getByTestID('label-card')
    .contains(oldLabelName)
    .should('be.visible')
  ...
  // modify
  cy.getByTestID('label-card')
    .contains(oldLabelName).click()
}
    \end{minted}
    \caption{{\tt influxdb} "Update Label" test code snippet. }
    \label{subfig:influxdb-rc-eg}
    \end{subfigure}
    \caption{{\tt influxdb} Network Resource Loading Example.}
    \label{fig:influxdb-rc-eg}
\end{figure}

Another example is found in the {\tt influxdb}~\cite{influxdb-rc-eg} project. This project provides a client side platform meant for storing, querying, and visualizing time series data. Figure~\ref{fig:influxdb-rc-eg} shows a flaky test for the label UI components. The test checks that labels update properly by first creating a new label and then attempting to modify the label's name and description through the UI. Figure~\ref{subfig:influxdb_screenshot_rc} presents the view of the test suite being run on the left with the UI view on the right. Figure~\ref{subfig:influxdb-rc-eg} shows the code snippet of the test corresponding to the screenshot. Lines 6-11 create the label to be used in the test. Lines 13-17 perform assertions on the labels retrieved through a network call. However, due to the execution timing, the label is not yet created in the backend store. The network call returns an empty response which causes the assertion on Line 15 to fail.

\paragraph {Resource Rendering}
Flaky tests in this category attempt to perform an action on a UI component before it is fully rendered. This attempt to interact with the missing component leads to visual differences detected by screenshot tests, or exceptions thrown by attempting to access elements that have not fully loaded yet. An example of this is seen in the {\tt generator-jhipster}\cite{generator-jhipster} project. This project provides a platform to generate modern web application with Java.
In this project, a test script attempts to click on a button and wait for the button to be displayed instead of the button being clickable. Normally, these descriptions refer to the same event, but the modal overlay shown in the UI can block the target button from being clickable. The faulty code snippet is shown in Figure~\ref{fig:jhipster-rc-eg}. The {\tt waitUntilDisplayable} function on Line 2 pauses the execution until the button is displayed on the page. The test can fail intermittently if another element is still above the button when Line 3 is reached, such as an element acting as a background shade in a confirmation modal.

\begin{figure}[ht]
\begin{minted}
[highlightlines={2}]
{js}
const modifiedDateSortButton = getModifiedDateSortButton();
await waitUntilDisplayed(modifiedDateSortButton);
await modifiedDateSortButton.click();
\end{minted}
\caption{{\tt generator-jhipster} Resource Rendering Example.}
\label{fig:jhipster-rc-eg}
\vspace{-8pt}
\end{figure}

This issue also appeared on the Android test, in the {\tt Volley}\cite{volley} project, the code snippet in Figure~\ref{fig:volley} leads to flaky behavior because of a short timeout. The listener occasionally does not finish executing in 100 ms timeout. This conflicts with the next request for verifying the order of calls.

\begin{figure}[hbpt]
\begin{minted}
[highlightlines={2,3}]
{java}
verifyNoMoreInteractions(listener);
verify(listener, timeout(100))
        .onRequestFinished(higherPriorityReq);
verify(listener, timeout(10))
        .onRequestFinished(lowerPriorityReq);
\end{minted}
\caption{{\tt Volley} Resource Loading Example.}
\label{fig:volley}
\end{figure}

\paragraph {Animation Timing Issue}
Flaky tests relying on animations are sensitive to timing differences in the running environment and may be heavily optimized to skip animation events. The sensitivity to scheduling in animations can lead to issues where assertions on the events are used to test for animation progress.

\begin{figure*}[hbpt]
    \centering
        \begin{subfigure}{0.3\textwidth}
            \centering
            \caption{Starting State}
            \label{subfig:plotly-rc-screnshots-1}
            \includegraphics
            [width=\textwidth]
            {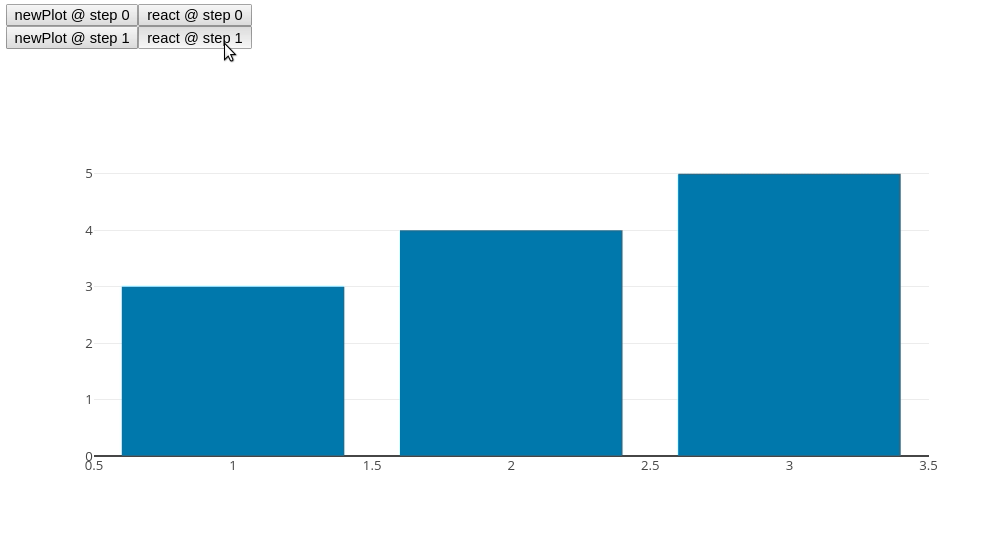}
        \end{subfigure}
        \begin{subfigure}{0.3\textwidth}
            \centering
            \caption{Background Axes Transition}
            \label{subfig:plotly-rc-screnshots-2}
            \includegraphics
            [width=\textwidth]
            {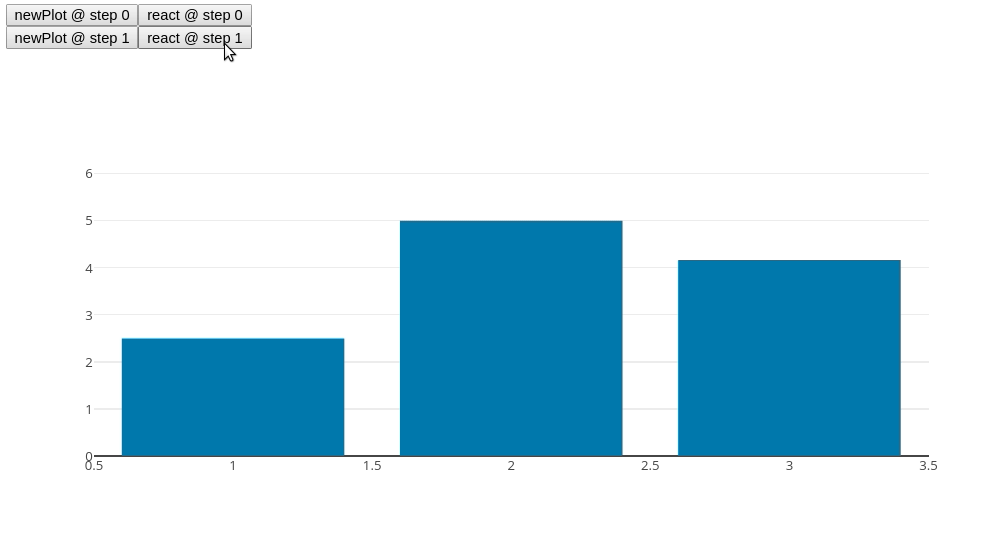}
        \end{subfigure}
        \begin{subfigure}{0.3\textwidth}
            \centering
            \caption{Bars Transition}
            \label{subfig:plotly-rc-screnshots-3}
            \includegraphics
            [width=\textwidth]
            {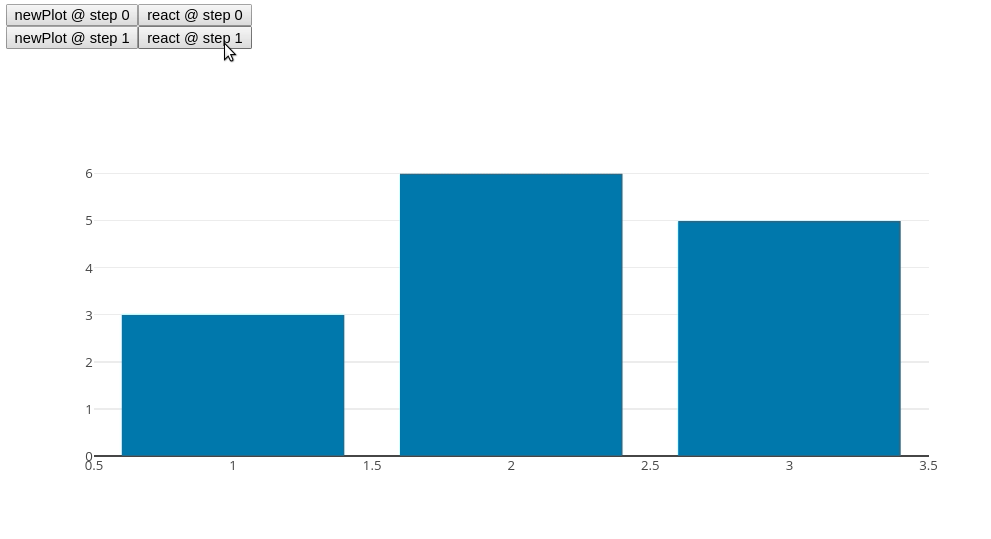}
        \end{subfigure}
    \caption{An animation timing issue in the {\tt plotly.js} project. (a) presents the initial state of a bar graph using the library. The bars start at values 3, 4, and 5, respectively. (b) The value of the bars are changed to 3, 6, and 5, respectively. The background axes change scale, but the bars are scaled incorrectly. (c) The bars then adjust to the correct scale. } 
    \label{fig:plotly-rc-screnshots}
\end{figure*}

An example of this type of issue is seen in the {\tt plotly.js} project~\cite{plotly.js}. This project provides visualization components such as bar graphs, line plots, and more for use in web pages. In the transition tests, the developers find that they intermittently fail due to an underlying race condition between the call to transition the axes and the call to transition the bar graphs. Depending on which transition is called first, assertions made on the layout of the graph may fail as the bar graph elements are in different positions than expected. In Figure~\ref{fig:plotly-rc-screnshots}, screenshots from a code snippet provided to reproduce the different states of the animation are shown. In Figure~\ref{subfig:plotly-rc-screnshots-1}, we see that graph starts with the first bar is on value 3, the second bar is on value 4, and the third bar is on value 5. Figure~\ref{subfig:plotly-rc-screnshots-2} shows the frame immediately after the “react @ step 1” button is clicked, changing the values of the bars to 3, 6, and 5 respectively. In this figure, the background lines of the axes have been shifted in order to represent the new scale, but the bars scale incorrectly to the new axes values. Finally, Figure~\ref{subfig:plotly-rc-screnshots-3}, the bars transition to their correct new values on the new axes. Since the bars are not in the expected positions during the transition test, the assertions made fail.

Another example is seen in the {\tt RXBinding}\cite{RxBinding} project for Android's UI widgets. In the {\tt RxSwipeRefreshLayoutTest}, which is used to test the swipe refresh gesture, the call to stop the refresh animation could happen anytime between the swipe release and the actual refresh animation. The behavior is flaky because the swipe animation timing used in the recorder is unable to catch up to the listener.

\subsubsection{Environment}
Some flaky tests manifest due to differences in the underlying platform used to run the tests. The platform can include the browser used for web projects and the version of Android, iOS, etc... used for mobile projects. We found that these issues can also be further divided into two subcategories.

\paragraph {Platform Issue}
These flaky tests suffer from an underlying issue in one particular platform that causes results obtained or actions performed to differ between consecutive runs within that same platform. In the {\tt ring-ui} project~\cite{ring-ui}, the screenshot tests for a drop-down component fail due to a rendering artifact bug present on Internet Explorer. This bug causes a slight variation around the drop-down border in the screenshots taken that cause the tests to fail when compared. These tests pass when run on other browsers.

One example on Android is about {\tt Androidx} navigation tool\cite{testing-samples}. For some versions of Android, Espresso has flaky behavior when performing a click action on the navigated page because sometimes it cannot close the navigation drawer before the click action. However, on other versions of Android, this test always passed.

\paragraph {Layout Difference}
Flaky tests can fail when the layout is different than what is expected due to differences in the browser environment. An example is found in the {\tt retail-ui} project~\cite{retail-ui-rc-eg}. This project contains a set of reusable components targeted at retail sites. The screenshot test for its dropdown component fails because different default window sizes across different browsers causes the dropdown box to be cut off in some browsers.

\subsubsection{Test Runner API Issue}
Another root cause of flakiness we found involved an issue when interacting with the APIs provided by the testing framework that caused it to function incorrectly. Flaky tests with this root cause either use the provided APIs incorrectly, or the flaky tests manage to expose an underlying issue in the provided API that causes the functionality to differ from what was expected. We also identify two subcategories among the flaky tests observed.

\paragraph {Incorrect Test Runner Interaction}

\begin{figure}[hbpt]
\centering
\includegraphics[width=0.5\linewidth]{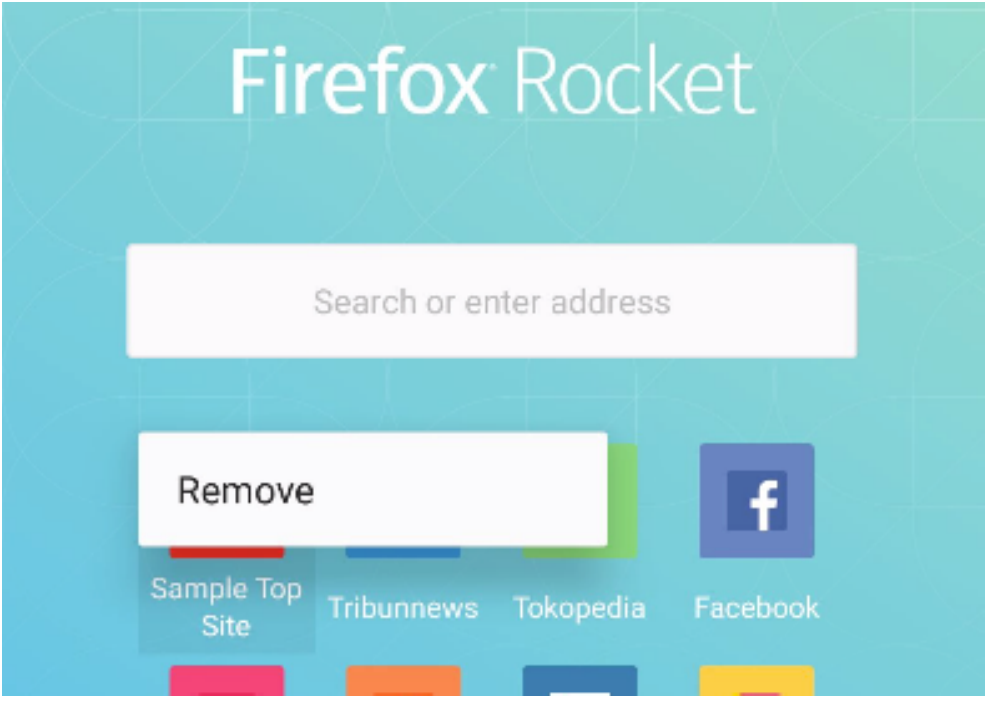}
\caption{{\tt FirefoxLite} Incorrect Test Runner Interaction Example.}
\label{fig:click}
\vspace{-8pt}
\end{figure}

UI tests use APIs provided by the test runner to interact with UI elements, but these APIs can hit unexpected behaviors that cause incorrect behavior.
For example, in the Android project {\tt FirefoxLite}\cite{mozilla-tw}, flakiness appeared because the testing model registered the click action by Espresso as a long click. Figure~\ref{fig:click} shows the UI layout after performing the click action incorrectly. A testing site should have opened by clicking “Sample Top Site” button. However, the “Remove” menu popped up instead because of the long click action on “Sample Top Site” button. This behavior difference caused the test to fail. 

\paragraph {DOM Selector Issue} 
Flaky tests interacting with DOM elements are intermittently unable to select the correct element due to differences in browser implementations or stale elements blocking focus. 
An example of the flakiness arising from an incorrect DOM element selection is found in the {\tt react-datepicker} project~\cite{react-datepicker-rc}. This project provides a reusable date picker component for the React library. The code under the test incorrectly sets two elements on the page to auto-focus on, causing a jump on the page that results in a visual flicker.

\subsubsection{Test Script Logic Issue}
In some flaky tests, flakiness arose due to incorrect logic within the test scripts. The flaky tests may have failed to clean data left by previous tests, made incorrect assertions during the test, loaded resources in an incorrect order, or incorrectly used a random data generator to create incompatible data. We find that tests in this category fall under one of four subcategories. 

\paragraph {Incorrect Resource Load Order}
Flaky tests in this category load resources after the calls that load the tests, causing the tested resources to be unavailable when the test is run. For example, in the project {\tt mapbox-navigation-android}\cite{mapbox}, the test crashed with an exception, because they duplicated a resource load call and then initialized a navigation activity.

\begin{figure}
    \centering
    \begin{subfigure}{\columnwidth}
        \includegraphics
        [width=\textwidth]
        {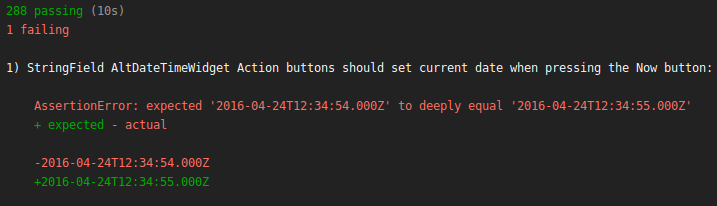}
        \caption{CI system failure in {\tt react-jsonschema-form} project when strictly comparing two date-time values. The values only differ by a marginal amount due to the time when the test is executed, but since the comparison is strict, the test will fail intermittently depending on when it is run.}
        \label{subfig:react-json-schema-rc-screenshot}
    \end{subfigure}
    
    \begin{subfigure}{\columnwidth}
        \begin{minted}
        [highlightlines={11-13}]
        {js}
it("should set current date when pressing the Now button", () => {
 const { node, onChange } = createFormComponent({
   schema: {
     type: "string",
     format: "date-time",
   },
   uiSchema,
 });

 Simulate.click(node.selector("a.btn-now"));
 const formValue = onChange.lastCall.args[0].formData;
 const expected = toDateString(parseDateString(new Date().toJSON(), true));
 expect(comp.state.formData).eql(expected);
});
        \end{minted}
        \caption{{\tt react-jsonschema-form} Strict Comparison Code snippet.}
        \label{subfig:react-jsonschema-rc-code}
    \end{subfigure}
    \caption{{\tt react-jsonschema-form} Strict Comparison Check Example.}
    \label{fig:react-json-schema-rc-eg}
    \vspace{-8pt}
\end{figure}

\paragraph {Time}
Flaky tests can fail when performing a comparison using a timestamp that may have changed from when it is created depending on the execution speed of the test.
An example is found in the 
{\tt react-jsonschema-form} project~\cite{react-jsonschema-form-fix-eg}. The project generates forms in React code by specifying the fields in JSON. In this project, a test on its date picker widget intermittently fails due to a strict comparison of time.
Figure~\ref{subfig:react-json-schema-rc-screenshot} shows a screenshot of a test failure within a CI system resulting from a strict comparison issue. Figure~\ref{subfig:react-jsonschema-rc-code} presents the faulty code snippet of the flaky test that intermittently fails in the CI system. Depending on when the test is run and how quickly the statements in the test execute, the date-time value retrieved on Line 11 with the date-time value generated in Line 12 can differ by a small amount, causing the assertion on Line 13 to fail. 

\paragraph {Test Order Dependency}
Flaky tests in this category can interfere with or be influenced by surrounding tests in the test suit. This interference can be done through shared data stores that are not cleaned well between test runs. As a result, the data stores may contain values from previous tests and produce incorrect values as a result. 
One example of this is appeared in Android project {\tt RESTMock}~\cite{restmock-103}. When trying to reset the server between tests, the test would sometimes return an exception, because there would be requests from the previous test still running as Android shares some processes between tests.

\paragraph {Randomness}
Tests can use random data generation, but these tests may intermittently fail for certain values of the data generated. An example of this type of failure is found in the {\tt nomad} project~\cite{nomad-rc-eg}, which provides a platform to deploy and manage containers. In this project, they find that tests utilizing the job factory component to generate fake tasks can intermittently fail when a job given a name or URL with spaces is created. This causes encoding issues later on in the tests. Since spaces are not valid in these fields, the spaces generated by the random string generator are edge cases that should have been handled.

\subsection{Results}
From these samples, we were able to find characteristics that are particular to flaky UI tests. The most predominant root cause for these flaky UI tests involved improper handling of asynchronous waiting mechanisms, such as the  mechanisms used when loading resources. These resources can include network resources as well as elements that have not yet been loaded in the page. This behavior resulted in erratic results in the tests, such as attempting to click buttons that had not yet opened. Many of these issues were resolved by refactoring the code to include delays when handling a potentially flaky call. We found that the root cause of the flaky behavior could present a challenge to find and properly fix, with some issues spanning over months to fix. In addition, the flaky nature led some of these issue reports to be closed and reopened in another report as many as five times. Other root causes included platform-specific behavior, layout differences, test order dependencies, and randomness. Platform-specific behavior produces flaky results for different runs in the same platform. Layout differences behavior causes flaky results due to inconsistencies across different platforms. Flakiness resulting from test order dependencies is caused by improper cleanup of data after runs of previous tests. UI tests involving random data generation can fail intermittently because of the characteristics of the data generated.

\section{Manifestation}
\label{sec:manifestation}
Reproducing flaky tests is a challenging task due to their inherit non-deterministic behavior. If developers provide details on how the flaky behavior was initially encountered and subsequently reproduced, this information provides possible strategies to apply to similar cases. We explore the strategies used by developers to manifest the underlying flaky behavior and construct categories for similar manifestations actions taken. These strategies are important when reporting the flaky test as they are inherently non-deterministic in nature, so it is challenging to reproduce them compared with regular bugs. Our categories are summarized in Table~\ref{table:manifestation-categories}.

\begin{table}[hbpt]
\caption{Summary of Manifestation Categories}
\label{table:manifestation-categories}
\centering
\def\arraystretch{0.5}
\resizebox{0.95\columnwidth}{!}{
    \begin{tabular}{@{}lrrr@{}}
    \toprule
    {\bf Manifestation Category}            & {\bf Web} & {\bf Mobile} & {\bf Total}\\ \midrule
    Unspecified                       & 101 & 40   &   141  \\ \midrule
    Specify Problematic Platform      & 21  & 17   &    38    \\ \midrule
    Reorder/Prune Test Suite          & 9   & 3    &    12    \\ \midrule
    Reset Configuration Between Tests & 2   & 7    &    9    \\ \midrule
    Provide Code Snippet              & 14  & 6    &    20    \\ \midrule
    Force Environment Conditions      & 5   & 10   &   15     \\ \midrule
    Total                            & 152 & 83   &   235     \\ \bottomrule
    \end{tabular}
}

\end{table}

\subsection{Specify Problematic Platform}
\begin{figure}[ht]
\centering
\includegraphics[width=0.95\linewidth]{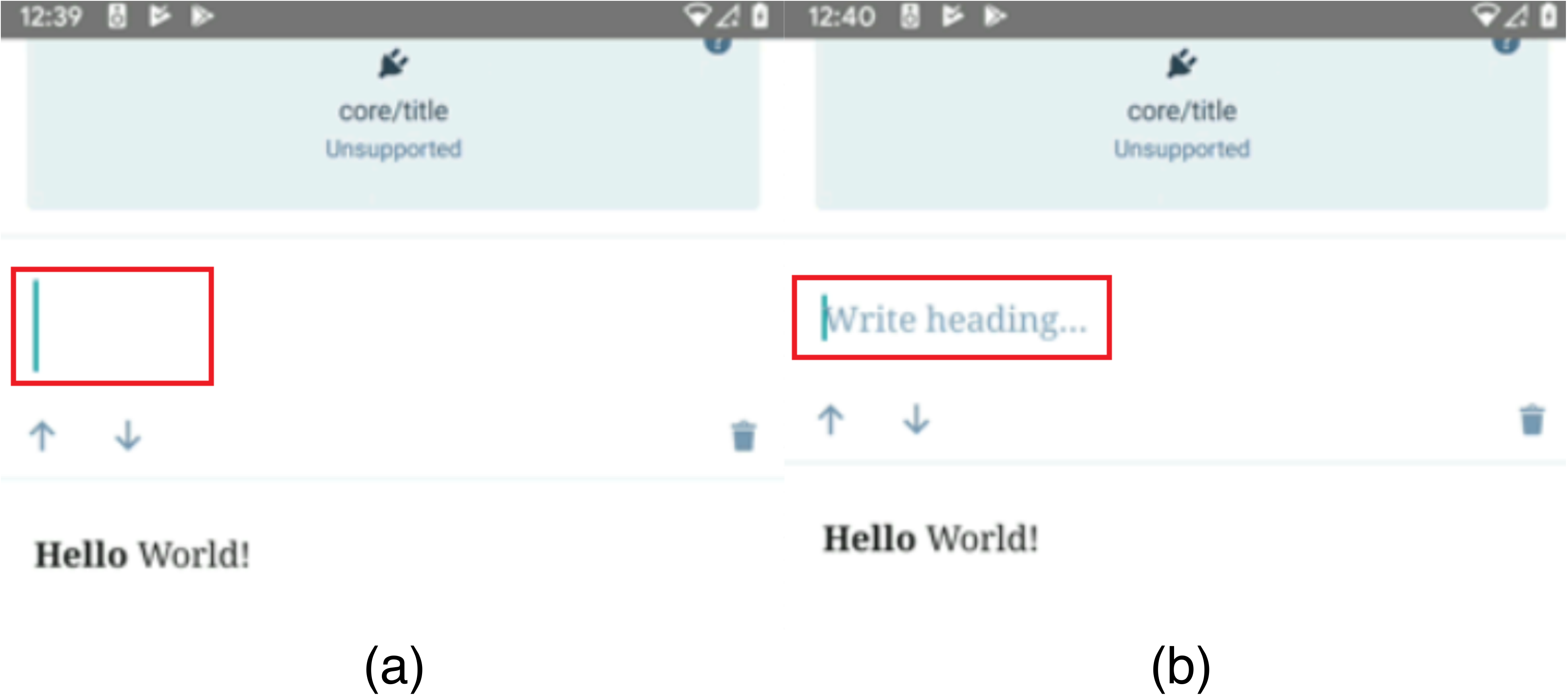}
\caption{{\tt Gutenberg-mobile} Specify Problematic Platform issue Example.}
\label{fig:type}
\end{figure}

Some tests are reported to only manifest on a specific platform. In this case, the author of the report specifies the problematic platform version to reproduce the flaky behavior. An example of this type of manifestation is found in the 
{\tt waterfox} project~\cite{waterfox-platform-eg}. This project is a web browser based on Firefox. In this project, an issue involving animation timing only manifests on MacOSX platforms. The report provides details on which file to run on this particular platform in order to reliably manifest the flaky behavior seen in the animation test. Another example in an Android project is from {\tt gutenberg-mobile}~\cite{gutenberg}. This project is the mobile version for Gutenberg Editor. Figure~\ref{fig:type} shows the bug that only appeared on the Google Pixel device with Android 10. When deleting the last character, the placeholder text should reveal as shown in Figure~\ref{fig:type}b; however, the text does not pop up. Instead, the screen appeared as shown in Figure~\ref{fig:type}a. Users would need to add an additional backspace key press to show the placeholder text.

\subsection{Reorder/Prune Test Suite}
Flakiness arising from test-order dependencies can be manifested by running the tests without the full test suite. This includes running tests by themselves, running tests in a different order, and changing the state between test runs in order to show the flaky behavior. An example of this manifestation strategy is seen in the {\tt influxdb} project~\cite{influxdb-test-suite-eg}. In this project, the flaky behavior surrounding table sorting is manifested by running the tests in the test suite independently. In some cases, trying to reset the environment configuration between tests can also lead to flakiness. In the project {\tt RESTMock}~\cite{restmock-103}, the developers tried to reduce flakiness by resetting the server configuration between tests. However, the test became more unstable because some Android processes were shared among these tests, and the forced reset caused concurrency conflicts.

\subsection{Provide Code Snippet}
Among the bug reports we observed, we find that some  reports include code snippets. The code snippets extract a portion of the flaky test into an enclosed sample to make reproducing the flaky behavior more reliable. An example of this strategy is used in the project {\tt plotly.js} project~\cite{plotly-manif-eg}. This projects provides data visualization components for use in web pages. In this project, a test for a treemap component contains a flaky image load. In order to manifest this more reliably, the reporter created a publicly-accessible code snippet that runs the component with the flaky loading behavior.

\subsection{Force Environment Conditions}
Flakiness that displays only when run on a specific platform or under certain environment settings can be manifested by forcibly setting these conditions, such as environment variables or browser window size, during the test run. An example of this can be found in the {\tt react-datepicker} project~\cite{react-datepicker-manif-eg}. This project provides a reusable datepicker component for use in React apps. A test for the calendar component has flaky behavior when run on the first two days of a new month. This behavior is manifested by setting the time used in the test to be one of these affected dates. Another example on Android is a click function in Espresso~\cite{espresso}. If we run an Espresso test which calls {\tt openActionBarOverflowOrOptionsMenu} on a slow device, a long-click action will be accidentally performed. This bug can be manifested by a short long-click timeout.

\section{Fixing Strategy}
\label{sec:fixing-strategies}
In this section, we examine the fixes of the flaky tests. We identify common fixing patterns and group them into categories. Through comparative analysis of root causes and fixing strategies, we find that most async wait issues are fixed by increasing delay or fixing the await mechanism used. The issues caused by the environments such as platform issues and layout differences normally could not be solved. The developers prefer to fix these tests by using a workaround or changing the library version. Table~\ref{table:fixing_category_stats} summarizes the categories and distribution of fixing strategies and are described in the following paragraphs.

\begin{table}[hbpt]
\caption{Summary of Fixing Categories Found}
\label{table:fixing_category_stats}
\def\arraystretch{1.2}
\resizebox{\linewidth}{!}{%
\begin{tabular}{@{}llrrr@{}}
\toprule
{\bf Categories} & {\bf Subcategories} & {\bf Web} & {\bf Mobile} & {\bf Total} \\ \midrule
\multirow{2}{*}{Delay} & Add/Increase Delay  &  14 & 7   & 21 \\
                       & Fix Await Mechanism &  35 & 8   & 43 \\ \hline
\multirow{2}{*}{Dependency} 
& Fix API Access   &   1 & 11    & 12 \\
& Change Library Version & 1 & 6  & 7 \\ \hline
Refactor Test & Refactor Logic Implementation & 49 & 26  & 75\\
\hline
Disable Features & Disable Animations  & 1  & 3    & 4  \\ \hline
Remove Test   & Remove Test         &  51  & 22   & 73  \\ \midrule
& Total  &  152  & 83  & 235 \\ \bottomrule
\end{tabular}
}
\end{table}

\subsection{Delay}
\subsubsection{Add or Increase Delay}
In order to reduce the chance of encountering flaky behavior, some tests will add or increase the delay between actions that involve fetching or loading. This prevents the rest of the test script from executing until the delay is up, giving the asynchronous call additional time to complete before moving on. An example of this fix is used in the {\tt next.js} project~\cite{next-js-fix-eg}. This project is used to generate complete web applications with React as the frontend framework. The patch increases the delays used in multiple steps as shown in Figure~\ref{fig:next-js-fix-eg}. In the figure, Line 1 loads a new browser instance and navigates to the “about” page. Lines 2 and 3 get the text on the page and assert that it is equal to the expected value. Lines 4 and 5 manipulate the about page's component file on the filesystem to make it invalid for use. Line 6 was the delay used before of 3 seconds. If the test is run during a heavy load on the CI, the operation in Line 5 may take longer than 3 seconds, so the fix is to update the wait to 10 seconds shown in Line 7. Finally, Line 9 makes the assertion that the updated text on the page shown matched the expected error message. 
While this does not fix the root cause directly, this code patch does decrease the chance of running into a timing issue during testing. 

\begin{figure}[ht]
\begin{minted}
[escapeinside=||]
{js}
const browser = await webdriver(context.appPort, '/hmr/about')
const text = await browser.elementByCss('p').text()
expect(text).toBe('This is the about page.')
const aboutPage = new File(join(__dirname, '../', 'pages', 'hmr', 'about.js'))
aboutPage.replace('export default', 'export default "not-a-page"\nexport const fn = ')
- await waitFor(3000)        
+ await waitFor(10000))    
expect(await browser.elementByCss('body').text())
.toMatch(/The default export is not a React Component/)
\end{minted}
\caption{{\tt next-js} Increase Delay Example.}
\label{fig:next-js-fix-eg}
\end{figure}

\subsubsection{Fix Waiting Mechanism}
In order to fix flaky behavior, some tests fix the mechanisms used to wait on an asynchronous call. This ensures that the call would finish before moving forward in the test script. An example is seen in the {\tt gestalt} project~\cite{gestalt-fix-eg}. This project contains a set of reusable components used on the Pinterest website. This test is run using a headless browser, and it is accessed through the {\tt page} variable. In 
Figure~\ref{fig:gestalt-fix-eg}, lines 2-10 emit an event on the page to trigger the action being tested. Line 12 is supposed to pause the script execution for 200 milliseconds in order for the page to complete the action from the event handler. However, the function {\tt page.waitFor} returns an asynchronous JavaScript promise, so it requires the {\tt await} keyword in order to allow the promise to resolve before the lines after the call are run. The issue is fixed by adding the {\tt await} keyword where needed.

\begin{figure}[ht]
\begin{minted}
[escapeinside=||]
{js}
it('removes all items', async () => {
 await page.evaluate(() => {
  window.dispatchEvent(
    new CustomEvent('set-masonry-items', {
      detail: {
        items: [],
      },
    })
  );
 });

-  page.waitFor(200);        
+  await page.waitFor(200); 

 const newItems = await page.$$(selectors.gridItem);
 assert.ok(!newItems || newItems.length === 0);
});
\end{minted}
\caption{{\tt Gestalt} Fix Waiting Example.}
\label{fig:gestalt-fix-eg}
\end{figure}

Another example in an Android project is from project {\tt RXBinding}\cite{RxBinding}. The developers avoided flakiness in this refresh layout test by manually removing the callbacks of {\tt stopRefreshing} and adding it back after 300 ms delay, if the motion {\tt ACTION\_UP} has been caught. The code snippet is shown in Figure~\ref{fig:rxbinding-fix}.

\begin{figure}[ht]
\begin{minted}
{java}
+ swipeRefreshLayout
+   .setId(R.id.swipe_refresh_layout);
+ swipeRefreshLayout.setOnTouchListener(new View.OnTouchListener() {
+   @Override public boolean onTouch(View v, MotionEvent event) {
+       if (MotionEventCompat.getActionMasked(event) == MotionEvent.ACTION_UP) {
+           handler.removeCallbacks(stopRefreshing);
+           handler.postDelayed(stopRefreshing, 300);
\end{minted}
\caption{{\tt RxBinding} Fix Waiting Mechanism Example.}
\label{fig:rxbinding-fix}
\end{figure}

\subsection{External Dependency}

\subsubsection{Fix Incorrect API Access}
Some tests resolved the flakiness by fixing the usage of an incorrect API function. After switching this function, the test script behaved as expected. An example is shown in the {\tt material-ui} project~\cite{material-ui-fix-eg}, which provides reusable web components implementing the Material design system. An API function from the testing library used to access DOM element children is incorrect. The code snippet in Figure~\ref{fig:material-ui-fix-eg} shows how the incorrect API function is fixed by calling the proper {\tt getPopperChildren} function instead of attempting to get the element's children directly. The correct function adds additional selection criteria in order to work within the template code generated by the third-party {\tt popper.js}~\cite{popper-js} framework.

\begin{figure}[ht]
\begin{minted}
{js}
- assert.strictEqual(
-  wrapper.find(Popper)
-  .childAt(0)
-  .hasClass(classes.tooltip),
-  true
- );

+ function getPopperChildren(wrapper) {
+  return new ShallowWrapper(
+   wrapper
+    .find(Popper)
+    .props()
+    .children({ popperProps: { style: {} }, restProps: {} }),
+   null
+  );
+ }

+ const popperChildren = getPopperChildren(wrapper);
+ assert.strictEqual(
+  popperChildren.childAt(0)
+  .hasClass(classes.tooltip),
+ true);
\end{minted}
\caption{{\tt material-ui} Fix Incorrect API Example.}
\label{fig:material-ui-fix-eg}
\end{figure}

Another example on the Android platform is found in the {\tt Detox} project\cite{detox}. The action to launch an application in an existing instance, which has launched an app during initialization, can lead to flaky behavior. Launching an app dynamically in UIAutomator is performed by moving to the recent-apps view and then selecting the app name. However, sometimes the recent-apps view shows the full activity name (e.g. com.wix.detox.MainActivity), instead of app name (e.g. Detox), which causes flakiness. To fix this bug, developer removed the UIAUtomator API and created new instances for each launch request. Figure~\ref{fig:detox-code} shows the code snippet of this fixing process.

\begin{figure}[]
\begin{minted}
{java}
- device.pressRecentApps();
- UiObject recentApp = device.findObject(selector .descriptionContains(appName));
- recentApp.click();

+ final Activity activity = ActivityTestRule.getActivity();
+ final Context appContext = activity.getApplicationContext();
+ final Intent intent = new Intent(appContext, activity.getClass());
+ intent.setFlags(Intent .FLAG_ACTIVITY_SINGLE_TOP);
+ launchActivitySync(intent);
\end{minted}
\caption{{\tt Detox} Fix Incorrect API Example.}
\label{fig:detox-code}
\end{figure}

\subsubsection{Change Library Version} %
Some tests changed the version of a dependency used in the test as the developers found
that the new version 
introduced the flaky behavior.

\subsection{Refactor Test Checks}
\subsubsection{Refactor Logic Implementation}
Some tests made changes to the logic used when performing checks in order to improve the intended purpose of the test while removing the flakiness observed in the test. An example is found in the 
{\tt react-jsonschema-form} project~\cite{react-jsonschema-form-fix-eg}. In the repository, a check between consecutive timestamps is given an additional error margin to handle the case of slow execution. Figure~\ref{fig:react-jsonschema-form-fix-eg} shows the code snippet changing the exact date-time comparison in Line 3 to the comparsion with an error margin of 5 seconds in Line 8.

\begin{figure}[ht]
\begin{minted}
{js}
- const expected = toDateString(
    parseDateString(new Date().toJSON(), true));
- expect(comp.state.formData).eql(expected);
+ // Test that the two DATETIMEs are within 5 seconds of each other.
+ const now = new Date().getTime();
+ const timeDiff = now - new Date(comp.state.formData)
    .getTime();
+ expect(timeDiff).to.be.at.most(5000);
\end{minted}
\caption{{\tt react-jsonschema-form} Refactor Logic Implementation Example.}
\label{fig:react-jsonschema-form-fix-eg}
\end{figure}

\subsection{Disable Features During Testing}
\subsubsection{Disable Animations}
In order to remove flakiness caused by animation timing, some test completely disabled animations during their run. This change removed the concern of ensuring an animation had completely finished before proceeding with the rest of the script.
An example of this is seen in the the {\tt wix-style-react} project where code is added to disable CSS animations when the test suite is run~\cite{wix-rc-eg}. Figure~\ref{fig:wix-style-react-fix-eg} shows the {\tt disableCSSAnimation} function defined on Lines 1-15 CSS rules disabling all transitions and animations. Line 21 adds a call to this function before all tests in the test suite are run.
\begin{figure}[ht]
\begin{minted}
{js}
+ export const disableCSSAnimation = () => {
+ const css = '* {' +
+ '-webkit-transition-duration: 0s !important;' +
+ 'transition-duration: 0s !important;' +
+ '-webkit-animation-duration: 0s !important;' +
+ 'animation-duration: 0s !important;' +
+ '}',
+ head = document.head || 
+         document.getElementsByTagName('head')[0],
+ style = document.createElement('style');

+ style.type = 'text/css';
+ style.appendChild(document.createNode(css));
+ head.appendChild(style);
+ };
...
beforeAll(() => {
  browser.get(storyUrl);
+ browser.executeScript(disableCSSAnimation);
});
\end{minted}
\caption{{\tt wix-style-react} Disable Animations Example.}
\label{fig:wix-style-react-fix-eg}
\end{figure}

\subsection{Removing Tests From Test Suite}
\subsubsection{Remove Tests}
In order to fix the test suite runs, some projects choose to remove these tests from the suite. This fix removes the flakiness in the test suite attributed to the flaky test being removed but reduces the code coverage.

\subsubsection{Mark Tests as Flaky}
Some tests are not entirely removed from the test suite. Instead, they are marked as being flaky which means that if the test fails, the entire test suite does not fail. This allows the test suite to be isolated from the effects of the flaky test without completely removing the coverage it provides.

\subsubsection{Blacklist Tests}
In order to conditionally prevent some tests from running, tests are added to a blacklist. The test in these blacklists can be skipped from test runs by setting the appropriate options for when the blacklist should be used.

\section{Discussion and Implications}
\label{sec:root-cause-fix-relationship}
\begin{figure}
    \centering
    \includegraphics[width=\columnwidth]{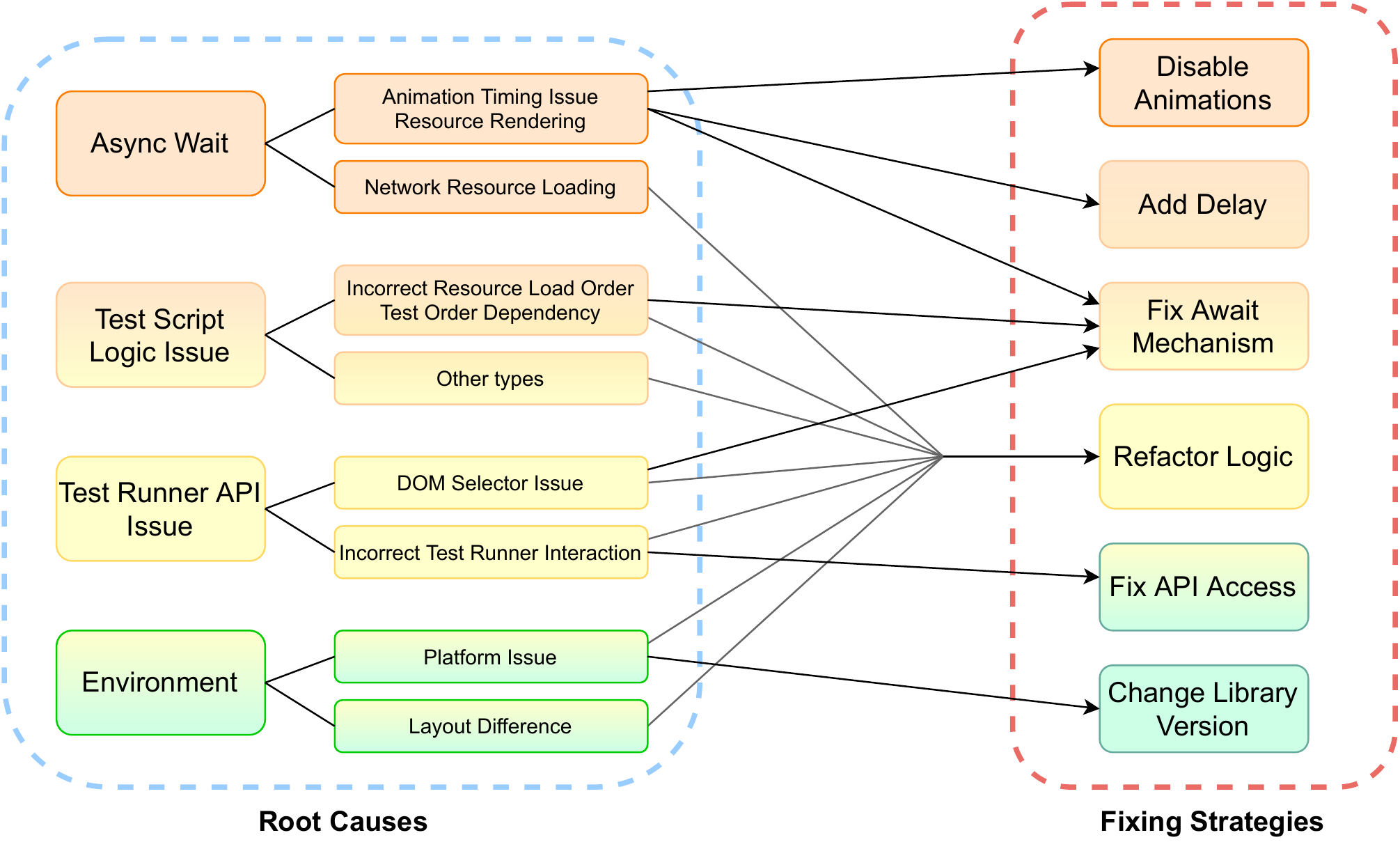}
    
    \caption{Relationship Between Root Causes and Fixing Strategies.}
    \label{fig:bug-map}
\end{figure}

We investigate our collected flaky UI tests to identify relationships between the root causes, manifestation strategies, and fixing strategies defined in Sections~\ref{sec:root-cause},~\ref{sec:manifestation}, and ~\ref{sec:fixing-strategies}, respectively.

Through our inspection, we can identify relationships between the underlying root causes in issues and how the issue was fixed. These relationships are presented in Figure~\ref{fig:bug-map}.

The goal of our study on flaky UI tests is to gain insights for designing automated flaky UI test detection and fixing approaches, so we analyze our dataset to identify correlations between manifestation strategies and root causes. However, we find that no strong correlations between these two groups exist in the dataset. Similarly, we could not identify strong correlations between manifestation strategy and fixing strategy. This leaves the question of detection strategies for flaky UI tests left open for future work to address. Our results do support relationships between root causes and fixing strategies. If the root cause of a flaky UI test is known, the relationships we draw in %
Figure~\ref{fig:bug-map} 
can be used to select an appropriate fixing strategy. 

Preliminary design ideas can be made for some of the fixing strategies we identify in Section~\ref{sec:fixing-strategies}. For the \textit{Add/Increase Delay} fixing strategy, a possible automated implementation could identify statements that perform the timing delay and increase the amount of time specified. If there is no delaying statement, then a delay can be after asynchronous function calls are performed. Granular details such as the amount of time to add in the delay or reliably identifying asynchronous function calls requires further analysis on the collected samples. Using the relationships found in  %
Figure~\ref{fig:bug-map}, 
this approach can be used to fix issues caused by \textit{Resource Rendering} (22.4\%) and \textit{Animation Timing Issue} (15.4\%).
For the \textit{Fix Await Mechanism} fixing strategy, an approach for automatic repair would be to identify statements that implement asynchronous wait mechanisms incorrectly. The details for this approach would be dependent on the language of the project and would require further analysis of the collected samples. This approach for an automated implementation of the \textit{Fix Await Mechanism} can be used to fix \textit{Incorrect Resource Load Order} (52.4\%), \textit{Animation Timing Issue} (38.5\%), \textit{Resource Rendering} (20.7\%), and \textit{DOM Selector Issue} (18.8\%).
The \textit{Disable Animations} fix can be implemented by configuring the test environment to disable animations globally when setting up. This approach can be used to fix issues caused by \textit{Animation Timing Issue} (7.7\%), \textit{Resource Rendering} (1.7\%), and \textit{DOM Selector Issue} (6.25\%).
The \textit{Change Library Version} fixing strategy could be automated by methodically switching different versions of the dependencies used in the project. This approach could be used to address issues caused by \textit{Animation Timing Issue} (7.7\%) and \textit{Test Runner API Issue} (6.25\%).

\section{Threats to Validity}
The results of our study are subject to several threats, including the representativeness of the projects inspected, the correctness of the methodology used, and the generalizability of our observations.

Regarding the representativeness of the projects in our dataset, we focused on the most popular repositories associated with popular frameworks on web and Android. We restrict the repositories to focus on repositories that impact real applications as opposed 
repositories under heavy development.
For mobile projects, we searched through GitHub database with strict and clear condition settings to ensure that the samples we obtain are targeted and representative. 

In respect to the correctness of the methodology used, we collect all available commits on GitHub from the repositories related to popular web UI frameworks. We also leveraged the GitHub Archive repository to find all issues related to Android UI frameworks. We filter out irrelevant commits and issues using keywords and then manually inspect the remaining commits and issues in order to verify the relevance to flaky UI tests. Each sample was inspected by at least two people in order to achieve consensus on the data collected.

Regarding the generalizability of the implications made, we selected flaky test samples from actual projects used in the wild. In addition, the samples do include large-scale industrial projects, such as the Angular framework itself. We limit numeric implications only to the dataset collected, and focus on qualitative implications made on the features of the test samples.

\section{Related Work}

\noindent
\textbf{Empirical Studies on Software Defects.} 
There have been several prior studies analyzing the fault-related characteristics of software systems \cite{Chillarege-icse1991, Chou-sosp2001, Lu2008LearningFM,  Sahoo-icse2010, Sullivan-ftcs1992, Thung-issre2012, Yin-fse2011, Yin-sosp2011, Zhang-issta2018, Aranda-icse2009, Gu-dsn2003, Sun-issta2016}. 
For example, in Lu {\it et al.}~\cite{Lu2008LearningFM} an empirical study was conducted on concurrency bugs. In Sahoo {\it et al.}~\cite{Sahoo-icse2010}, bugs in server software were studied, and in Chou {\it et al.} \cite{Chou-sosp2001}, operating system errors were investigated. 

\noindent
\textbf{Studying Flaky Tests.} Flaky tests have gained interest among the academic community. These tests were first looked at in 2014 by Luo {\it et al.}~\cite{luo_2014}. In this study, 201 commits from 51 open-source Java projects were manually inspected and categorized into 11 categories. Later, Zhang {\it et al.} (2014)~\cite{zhang-2014} performed studied flaky tests specifically caused by test order dependencies. In 2015, Goa {\it et al.}~\cite{gao-2015} conducted a study that concluded that reproducing flaky tests can be difficult. 
Thorve {\it et al.} (2018)~\cite{thorve-2018} studied 29 Android projects with 77 commits related to flakiness. They found three new categories differing from the ones identified in earlier studies: Dependency, Program Logic, and UI.
Lam {\it et al.} (2019)~\cite{lam2019root} examine the presence of flaky test in large-scale industrial projects and find that flaky test cause a significant impact on build failure rates.
Morán {\it et al.} (2019)~\cite{moran-2019} develop the FlakcLoc technique to find flaky tests in web applications by executing them under different environment conditions.
Eck {\it et al.} (2019)~\cite{flaky_tests_mozilla_2019} survey 21 professional developers from Mozilla to learn about the perceptions that developers have on the impacts that flaky tests cause during development.
Dong {\it et al.} (2020)~\cite{dong2020concurrencyrelated} inspect 28 popular Android apps and 245 identified flaky tests to develop their FlakeShovel technique that controls and manipulates thread execution.
Lam {\it et al.} (2020)~\cite{lam2020} study the lifecycle of flaky tests in large-scale projects at Microsoft by focusing on the timing between flakiness reappearance, the runtime of the tests, and the time to fix the flakiness. 

\noindent
\textbf{Detecting and Fixing Flaky Tests.}
Bell {\it et al.} (2018)~\cite{bell2018deflaker} developed the technique DeFlaker to detect flaky tests by monitoring the coverage of code changes in the executing build with the location that triggered the test failure. Flaky tests were those that failed without executing any of the new code changes. 
Lam {\it et al.} (2019)~\cite{lam2019root} develop the framework RootFinder to identify flaky tests and their root causes through dynamic analysis. The tool iDFlakies can detect flaky tests and classify the tests into order-dependent and non-order-dependent categories~\cite{lamIDFlakiesFrameworkDetecting2019}.
Shi {\it et al.} (2019)~\cite{shi-2019} develop the tool iFixFlakies to detect and automatically fix order-dependent tests by using code from other tests within a test suite to suggest a patch. 
Terragni {\it et al.} (2020)~\cite{terragni-2020} proposed a technique to run flaky tests in multiple containers with different environments simultaneously.

\section{Conclusions}
This paper performs a study on flakiness arising in UI tests in both web and mobile projects. We investigated 235 flaky tests collected from 25 web and 37 mobile popular GitHub repositories. The flaky test samples are analyzed to identify the typical root causes of the flaky behavior, the manifestation strategies used to report and reproduce the flakiness, and the common fixing strategies applied to these tests to reduce the flaky behavior. Through our analysis, we present findings on the prevalence of certain root causes, the differences that root causes appear between web and mobile platforms, and the differences in the rates of fixing strategies applied. We believe our analysis can provide guidance towards developing effective detection and prevention techniques specifically geared towards flaky UI tests. We make our dataset available at \href{https://ui-flaky-test.github.io/}{https://ui-flaky-test.github.io/}.
\section{Acknowledgments}

We thank the anonymous reviewers for their constructive comments. This research was partially supported by NSF 2047980 and Facebook Testing and Verification Research Award (2019). Any opinions, findings, and conclusions in this paper are those of the authors only and do not necessarily reflect the views of our sponsors.

\bibliographystyle{IEEEtran}
\bibliography{bibliography}

\begin{thebibliography}{10}
\providecommand{\url}[1]{#1}
\csname url@samestyle\endcsname
\providecommand{\newblock}{\relax}
\providecommand{\bibinfo}[2]{#2}
\providecommand{\BIBentrySTDinterwordspacing}{\spaceskip=0pt\relax}
\providecommand{\BIBentryALTinterwordstretchfactor}{4}
\providecommand{\BIBentryALTinterwordspacing}{\spaceskip=\fontdimen2\font plus
\BIBentryALTinterwordstretchfactor\fontdimen3\font minus
  \fontdimen4\font\relax}
\providecommand{\BIBforeignlanguage}[2]{{%
\expandafter\ifx\csname l@#1\endcsname\relax
\typeout{** WARNING: IEEEtran.bst: No hyphenation pattern has been}%
\typeout{** loaded for the language `#1'. Using the pattern for}%
\typeout{** the default language instead.}%
\else
\language=\csname l@#1\endcsname
\fi
#2}}
\providecommand{\BIBdecl}{\relax}
\BIBdecl

\bibitem{micco2017state}
J.~Micco, ``The state of continuous integration testing@ google,'' 2017.

\bibitem{luo_2014}
\BIBentryALTinterwordspacing
Q.~Luo, F.~Hariri, L.~Eloussi, and D.~Marinov, ``An empirical analysis of flaky
  tests,'' in \emph{Proceedings of the 22nd ACM SIGSOFT International Symposium
  on Foundations of Software Engineering}, ser. FSE 2014.\hskip 1em plus 0.5em
  minus 0.4em\relax New York, NY, USA: Association for Computing Machinery,
  2014, p. 643–653. [Online]. Available:
  \url{https://doi.org/10.1145/2635868.2635920}
\BIBentrySTDinterwordspacing

\bibitem{bell2018deflaker}
J.~Bell, O.~Legunsen, M.~Hilton, L.~Eloussi, T.~Yung, and D.~Marinov,
  ``Deflaker: Automatically detecting flaky tests,'' in \emph{2018 IEEE/ACM
  40th International Conference on Software Engineering (ICSE)}.\hskip 1em plus
  0.5em minus 0.4em\relax IEEE, 2018, pp. 433--444.

\bibitem{lam2019root}
W.~Lam, P.~Godefroid, S.~Nath, A.~Santhiar, and S.~Thummalapenta, ``Root
  causing flaky tests in a large-scale industrial setting,'' in
  \emph{Proceedings of the 28th ACM SIGSOFT International Symposium on Software
  Testing and Analysis}, 2019, pp. 101--111.

\bibitem{github-search-api}
\BIBentryALTinterwordspacing
``Search,'' 2020. [Online]. Available:
  \url{https://docs.github.com/en/rest/reference/search}
\BIBentrySTDinterwordspacing

\bibitem{gharchive}
\BIBentryALTinterwordspacing
Github, ``Github archive.'' [Online]. Available:
  \url{https://archiveprogram.github.com/}
\BIBentrySTDinterwordspacing

\bibitem{ring-ui-rc-eg}
\BIBentryALTinterwordspacing
``chore: use base64 uri decoded avatar to avoid flaky ui tests if image,''
  2020. [Online]. Available:
  \url{https://github.com/JetBrains/ring-ui/commit/f7bc28af06433ff22e898aacd2b3e8f0534defda}
\BIBentrySTDinterwordspacing

\bibitem{influxdb-rc-eg}
\BIBentryALTinterwordspacing
``fix(e2e): fix race conditions,'' 2019. [Online]. Available:
  \url{https://github.com/influxdata/influxdb/commit/4f5ff962d69a84f7a6970b02f9e79b09dbad21fe}
\BIBentrySTDinterwordspacing

\bibitem{generator-jhipster}
pascalgrimaud, ``react: Fix intermittent e2e failures,''
  \url{https://github.com/jhipster/generator-jhipster/commit/2865e441e4b09335f88f3839ee9147f8b8b9c05e},
  2019.

\bibitem{volley}
sphill99 and S.~Phillips, ``google/volley,'' 2017.

\bibitem{plotly.js}
alexcjohnson, ``one more flaky test suite,''
  \url{https://github.com/plotly/plotly.js/commit/a2fc07a187c4d26bf2f1bcb3e2aa806b75ad24fc},
  2018.

\bibitem{RxBinding}
nojunpark, ``Fix rxswipedismissbehavior flaky test,''
  \url{https://github.com/JakeWharton/RxBinding/commit/affa7a4f58e5becec4ad8b49d30f525d6ad4c2a6},
  2016.

\bibitem{ring-ui}
princed, ``Concurrent modification exception,''
  \url{https://github.com/JetBrains/ring-ui/commit/5d9f96d6ffa3a3c99722047677d5a545c02bdd80},
  2017.

\bibitem{testing-samples}
\BIBentryALTinterwordspacing
chklow, ``Espresso is not waiting for drawer to close,'' 2019. [Online].
  Available: \url{https://github.com/android/testing-samples/issues/289}
\BIBentrySTDinterwordspacing

\bibitem{retail-ui-rc-eg}
\BIBentryALTinterwordspacing
``test(dropdowncontainer): fix flaky screenshot test,'' 2019. [Online].
  Available:
  \url{https://github.com/skbkontur/retail-ui/commit/a006fdf0e0e65d5fde07134c6909870666e7947f}
\BIBentrySTDinterwordspacing

\bibitem{mozilla-tw}
\BIBentryALTinterwordspacing
cnevinc, ``[ui test intermittent],'' 2018. [Online]. Available:
  \url{https://github.com/mozilla-tw/FirefoxLite/issues/2549}
\BIBentrySTDinterwordspacing

\bibitem{react-datepicker-rc}
\BIBentryALTinterwordspacing
Hacker0x01, ``Remove second autofocus example,'' 2018. [Online]. Available:
  \url{https://github.com/Hacker0x01/react-datepicker/pull/1390/commits/8fc31964251944be79f2e8699b79e5f39080272f}
\BIBentrySTDinterwordspacing

\bibitem{mapbox}
Guardiola31337, ``Flaky navigationvieworientationtest,''
  \url{https://github.com/mapbox/mapbox-navigation-android/issues/1209}, 2018.

\bibitem{react-jsonschema-form-fix-eg}
\BIBentryALTinterwordspacing
``merge pull request \#167 from mozilla-services/162-fix-intermittent-da,''
  2020. [Online]. Available:
  \url{https://github.com/rjsf-team/react-jsonschema-form/commit/2318786b38ead5eddc7c0e3146825f19013e0beb}
\BIBentrySTDinterwordspacing

\bibitem{restmock-103}
\BIBentryALTinterwordspacing
Sloy, ``Concurrent modification exception,'' 2019. [Online]. Available:
  \url{https://github.com/andrzejchm/RESTMock/issues/103}
\BIBentrySTDinterwordspacing

\bibitem{nomad-rc-eg}
\BIBentryALTinterwordspacing
``Ui: Fix a couple flaky tests,'' 2018. [Online]. Available:
  \url{https://github.com/hashicorp/nomad/pull/4167/commits/69251628f7a3f03ce603abfea5c8f48b4804c39e}
\BIBentrySTDinterwordspacing

\bibitem{waterfox-platform-eg}
\BIBentryALTinterwordspacing
MrAlex94, ``Bug 1504929 - start animations once after a mozreftestinvalidate,''
  2018. [Online]. Available:
  \url{https://github.com/MrAlex94/Waterfox/commit/23793e3a2172787eca440889a8c4ec3cc6069862}
\BIBentrySTDinterwordspacing

\bibitem{gutenberg}
\BIBentryALTinterwordspacing
mchowning, ``Deleting heading block content requires extra backspace to show
  placeholder,'' 2020. [Online]. Available:
  \url{https://github.com/wordpress-mobile/gutenberg-mobile/issues/1663}
\BIBentrySTDinterwordspacing

\bibitem{influxdb-test-suite-eg}
\BIBentryALTinterwordspacing
influxdata, ``fix(ui): front end sorting for numeric values now being
  handled,'' 2019. [Online]. Available:
  \url{https://github.com/influxdata/influxdb/commit/bba04e20b44dd0f8fd049d80f270424eb266533f}
\BIBentrySTDinterwordspacing

\bibitem{plotly-manif-eg}
\BIBentryALTinterwordspacing
etpinard, ``add treemap\_coffee to list of flaky image tests,'' 2020. [Online].
  Available:
  \url{https://github.com/plotly/plotly.js/commit/66156054cb08b90bc50219ff9a2baeebb674c580}
\BIBentrySTDinterwordspacing

\bibitem{react-datepicker-manif-eg}
\BIBentryALTinterwordspacing
aij, ``Fix flaky failing test,'' 2017. [Online]. Available:
  \url{https://github.com/Hacker0x01/react-datepicker/commit/db64f070d72ff0705239f613bd5bba9602d3742f}
\BIBentrySTDinterwordspacing

\bibitem{espresso}
\BIBentryALTinterwordspacing
``Espresso,'' 2020. [Online]. Available:
  \url{https://developer.android.com/training/testing/espresso}
\BIBentrySTDinterwordspacing

\bibitem{next-js-fix-eg}
\BIBentryALTinterwordspacing
vercel, ``introduce dynamic(() => import()),'' 2020. [Online]. Available:
  \url{https://github.com/vercel/next.js/commit/42736c061ad0e5610522de2517c928b2b8af0ed4}
\BIBentrySTDinterwordspacing

\bibitem{gestalt-fix-eg}
\BIBentryALTinterwordspacing
pinterest, ``masonry: masonryinfinite for infinite fetching (\#307),'' 2020.
  [Online]. Available:
  \url{https://github.com/pinterest/gestalt/commit/f6c683b66b2d8b0ec87db283418459e87160a21f}
\BIBentrySTDinterwordspacing

\bibitem{material-ui-fix-eg}
\BIBentryALTinterwordspacing
``[test] fix flaky popper.js test,'' 2020. [Online]. Available:
  \url{https://github.com/mui-org/material-ui/commit/9d1c2f0ab014c76ddc042dea58a6a9384fc108f4}
\BIBentrySTDinterwordspacing

\bibitem{popper-js}
\BIBentryALTinterwordspacing
popperjs, ``Tooltip \& popover positioning engine,'' 2020. [Online]. Available:
  \url{https://github.com/popperjs/popper-core}
\BIBentrySTDinterwordspacing

\bibitem{detox}
\BIBentryALTinterwordspacing
d4vidi, ``Fix consecutive app-launches issue,'' 2019. [Online]. Available:
  \url{https://github.com/wix/Detox/pull/1690/commits/c982798e8904b8384e4966f4ed20700b66921b399}
\BIBentrySTDinterwordspacing

\bibitem{wix-rc-eg}
\BIBentryALTinterwordspacing
``skip flaky visual eyes test (\#3306),'' 2020. [Online]. Available:
  \url{https://github.com/wix/wix-style-react/commit/ddebb9fc31f3aaea7b80dea034c3baa256ec2b74}
\BIBentrySTDinterwordspacing

\bibitem{Chillarege-icse1991}
R.~{Chillarege}, W.~. {Kao}, and R.~G. {Condit}, ``Defect type and its impact
  on the growth curve (software development),'' in \emph{[1991 Proceedings]
  13th International Conference on Software Engineering}, 1991, pp. 246--255.

\bibitem{Chou-sosp2001}
A.~Chou, J.~Yang, B.~Chelf, S.~Hallem, and D.~Engler, ``An empirical study of
  operating systems errors,'' \emph{Operating Systems Review (ACM)}, vol.~35,
  09 2001.

\bibitem{Lu2008LearningFM}
S.~Lu, S.~Park, E.~Seo, and Y.~Zhou, ``Learning from mistakes: a comprehensive
  study on real world concurrency bug characteristics,'' in \emph{ASPLOS},
  2008.

\bibitem{Sahoo-icse2010}
\BIBentryALTinterwordspacing
S.~K. Sahoo, J.~Criswell, and V.~Adve, ``An empirical study of reported bugs in
  server software with implications for automated bug diagnosis,'' in
  \emph{Proceedings of the 32nd ACM/IEEE International Conference on Software
  Engineering - Volume 1}, ser. ICSE ’10.\hskip 1em plus 0.5em minus
  0.4em\relax New York, NY, USA: Association for Computing Machinery, 2010, p.
  485–494. [Online]. Available: \url{https://doi.org/10.1145/1806799.1806870}
\BIBentrySTDinterwordspacing

\bibitem{Sullivan-ftcs1992}
M.~{Sullivan} and R.~{Chillarege}, ``A comparison of software defects in
  database management systems and operating systems,'' in \emph{[1992] Digest
  of Papers. FTCS-22: The Twenty-Second International Symposium on
  Fault-Tolerant Computing}, 1992, pp. 475--484.

\bibitem{Thung-issre2012}
F.~{Thung}, S.~{Wang}, D.~{Lo}, and L.~{Jiang}, ``An empirical study of bugs in
  machine learning systems,'' in \emph{2012 IEEE 23rd International Symposium
  on Software Reliability Engineering}, 2012, pp. 271--280.

\bibitem{Yin-fse2011}
\BIBentryALTinterwordspacing
Z.~Yin, D.~Yuan, Y.~Zhou, S.~Pasupathy, and L.~Bairavasundaram, ``How do fixes
  become bugs?'' in \emph{Proceedings of the 19th ACM SIGSOFT Symposium and the
  13th European Conference on Foundations of Software Engineering}, ser.
  ESEC/FSE ’11.\hskip 1em plus 0.5em minus 0.4em\relax New York, NY, USA:
  Association for Computing Machinery, 2011, p. 26–36. [Online]. Available:
  \url{https://doi.org/10.1145/2025113.2025121}
\BIBentrySTDinterwordspacing

\bibitem{Yin-sosp2011}
\BIBentryALTinterwordspacing
Z.~Yin, X.~Ma, J.~Zheng, Y.~Zhou, L.~N. Bairavasundaram, and S.~Pasupathy, ``An
  empirical study on configuration errors in commercial and open source
  systems,'' in \emph{Proceedings of the Twenty-Third ACM Symposium on
  Operating Systems Principles}, ser. SOSP ’11.\hskip 1em plus 0.5em minus
  0.4em\relax New York, NY, USA: Association for Computing Machinery, 2011, p.
  159–172. [Online]. Available: \url{https://doi.org/10.1145/2043556.2043572}
\BIBentrySTDinterwordspacing

\bibitem{Zhang-issta2018}
\BIBentryALTinterwordspacing
Y.~Zhang, Y.~Chen, S.-C. Cheung, Y.~Xiong, and L.~Zhang, ``An empirical study
  on tensorflow program bugs,'' in \emph{Proceedings of the 27th ACM SIGSOFT
  International Symposium on Software Testing and Analysis}, ser. ISSTA
  2018.\hskip 1em plus 0.5em minus 0.4em\relax New York, NY, USA: Association
  for Computing Machinery, 2018, p. 129–140. [Online]. Available:
  \url{https://doi.org/10.1145/3213846.3213866}
\BIBentrySTDinterwordspacing

\bibitem{Aranda-icse2009}
J.~{Aranda} and G.~{Venolia}, ``The secret life of bugs: Going past the errors
  and omissions in software repositories,'' in \emph{2009 IEEE 31st
  International Conference on Software Engineering}, 2009, pp. 298--308.

\bibitem{Gu-dsn2003}
{Weining Gu}, Z.~{Kalbarczyk}, {Ravishankar}, K.~{Iyer}, and {Zhenyu Yang},
  ``Characterization of linux kernel behavior under errors,'' in \emph{2003
  International Conference on Dependable Systems and Networks, 2003.
  Proceedings.}, 2003, pp. 459--468.

\bibitem{Sun-issta2016}
\BIBentryALTinterwordspacing
C.~Sun, V.~Le, Q.~Zhang, and Z.~Su, ``Toward understanding compiler bugs in gcc
  and llvm,'' in \emph{Proceedings of the 25th International Symposium on
  Software Testing and Analysis}, ser. ISSTA 2016.\hskip 1em plus 0.5em minus
  0.4em\relax New York, NY, USA: Association for Computing Machinery, 2016, p.
  294–305. [Online]. Available: \url{https://doi.org/10.1145/2931037.2931074}
\BIBentrySTDinterwordspacing

\bibitem{zhang-2014}
\BIBentryALTinterwordspacing
S.~Zhang, D.~Jalali, J.~Wuttke, K.~Mu\c{s}lu, W.~Lam, M.~D. Ernst, and
  D.~Notkin, ``Empirically revisiting the test independence assumption,'' in
  \emph{Proceedings of the 2014 International Symposium on Software Testing and
  Analysis}, ser. ISSTA 2014.\hskip 1em plus 0.5em minus 0.4em\relax New York,
  NY, USA: Association for Computing Machinery, 2014, p. 385–396. [Online].
  Available: \url{https://doi.org/10.1145/2610384.2610404}
\BIBentrySTDinterwordspacing

\bibitem{gao-2015}
Z.~{Gao}, Y.~{Liang}, M.~B. {Cohen}, A.~M. {Memon}, and Z.~{Wang}, ``Making
  system user interactive tests repeatable: When and what should we control?''
  in \emph{2015 IEEE/ACM 37th IEEE International Conference on Software
  Engineering}, vol.~1, 2015, pp. 55--65.

\bibitem{thorve-2018}
S.~{Thorve}, C.~{Sreshtha}, and N.~{Meng}, ``An empirical study of flaky tests
  in android apps,'' in \emph{2018 IEEE International Conference on Software
  Maintenance and Evolution (ICSME)}, 2018, pp. 534--538.

\bibitem{moran-2019}
J.~Morán, C.~Augusto~Alonso, A.~Bertolino, C.~de~la Riva, and J.~Tuya,
  ``Debugging flaky tests on web applications,'' 01 2019, pp. 454--461.

\bibitem{flaky_tests_mozilla_2019}
\BIBentryALTinterwordspacing
M.~Eck, F.~Palomba, M.~Castelluccio, and A.~Bacchelli, ``Understanding flaky
  tests: The developer’s perspective,'' in \emph{Proceedings of the 2019 27th
  ACM Joint Meeting on European Software Engineering Conference and Symposium
  on the Foundations of Software Engineering}, ser. ESEC/FSE 2019.\hskip 1em
  plus 0.5em minus 0.4em\relax New York, NY, USA: Association for Computing
  Machinery, 2019, p. 830–840. [Online]. Available:
  \url{https://doi.org/10.1145/3338906.3338945}
\BIBentrySTDinterwordspacing

\bibitem{dong2020concurrencyrelated}
Z.~Dong, A.~Tiwari, X.~L. Yu, and A.~Roychoudhury, ``Concurrency-related flaky
  test detection in android apps,'' 2020.

\bibitem{lam2020}
\BIBentryALTinterwordspacing
W.~Lam, K.~Mu\c{s}lu, H.~Sajnani, and S.~Thummalapenta, ``A study on the
  lifecycle of flaky tests,'' in \emph{Proceedings of the ACM/IEEE 42nd
  International Conference on Software Engineering}, ser. ICSE '20.\hskip 1em
  plus 0.5em minus 0.4em\relax New York, NY, USA: Association for Computing
  Machinery, 2020, p. 1471–1482. [Online]. Available:
  \url{https://doi.org/10.1145/3377811.3381749}
\BIBentrySTDinterwordspacing

\bibitem{lamIDFlakiesFrameworkDetecting2019}
W.~Lam, R.~Oei, A.~Shi, D.~Marinov, and T.~Xie, ``{iDFlakies}: {A} {Framework}
  for {Detecting} and {Partially} {Classifying} {Flaky} {Tests},'' in
  \emph{2019 12th {IEEE} {Conference} on {Software} {Testing}, {Validation} and
  {Verification} ({ICST})}, Apr. 2019, pp. 312--322, iSSN: 2159-4848.

\bibitem{shi-2019}
\BIBentryALTinterwordspacing
A.~Shi, W.~Lam, R.~Oei, T.~Xie, and D.~Marinov, ``ifixflakies: A framework for
  automatically fixing order-dependent flaky tests,'' in \emph{Proceedings of
  the 2019 27th ACM Joint Meeting on European Software Engineering Conference
  and Symposium on the Foundations of Software Engineering}, ser. ESEC/FSE
  2019.\hskip 1em plus 0.5em minus 0.4em\relax New York, NY, USA: Association
  for Computing Machinery, 2019, p. 545–555. [Online]. Available:
  \url{https://doi.org/10.1145/3338906.3338925}
\BIBentrySTDinterwordspacing

\bibitem{terragni-2020}
F.~F. Valerio~Terragni, Pasquale~Salza, ``A container-based infrastructure for
  fuzzy-driven root causing of flaky tests,'' 2020.

\end{thebibliography}
\end{document}